# A real-time hole depth diagnostic based on coherent imaging with plasma amendment during femtosecond laser hole-drilling

P. Xu[1], Y. Yu[2,*], C.J. Xiao[3], R.J. Liu[1], K. Zha[1], L. Zhou[1], Y.T. Liu[1], Z. Xu[1]

[1] *School of Physics, Xihua University, Chengdu 610039, China*

[2] *Sino-French Institute of Nuclear Engineering and Technology, Sun Yat-sen University, Zhuhai 519082, China*

[3] *State Key Laboratory of Nuclear Physics and Technology, School of Physics, Peking University, Beijing 100871, China*

*Corresponding author：yuyi56@mail.sysu.edu.cn

**Abstract**

An in-process coherent imaging diagnostic has been developed to real-time measure the hole depth during air-film hole drilling by a femtosecond laser. A super-luminescent diode with a wavelength of $830 \pm 13$ nm is chosen as the coherent light source which determines a depth resolution of 12 μm. The drilled hole is coupled as a part of the sample arm and the depth variation can be extracted from the length variation of the optical path. Interference is realized in the detection part and a code has been written to discriminate the interference fringes. Density of plasma in the hole is diagnosed to evaluate its amendment to the optical path length and the depth measurement error induced by plasma is non-ignorable when drilling deep holes.

## 1. INTRODUCTION

As the great progress of modern industry, higher and higher requirements on precision machinery processing are advanced. Electrical discharge machining (EDM) [1] has worked as the main processing method for decades and achieves its great success in hole drilling in kinds of metals. Other techniques, such as electrochemical machining (ECM) [2], e-beam processing [3], ultrasonic processing [4] and laser drilling, have also been developed and applied in different cases. Early laser processing method uses long-pulse or steady-state high-energy laser beams to melt the metal sample and uses high-pressure air to help blowing the molten grains out of the hole. This method will probably lead to terrible recast layer in the holes [5,6]. In the engine component processing, recast layer results in great performance degradation of the air-film holes since it may contain oxide inclusions and microcracks. This hot processing will also cause local deformation of the metal and great laser energy wasting which reduces the processing efficiency. Femtosecond laser has been proven to be valuable to effectively solve these defections, although the physics of the interaction between the femtosecond laser pulses and the metal material is still not fully understood. When a laser pulse front reaches the metal surface, the free electrons will absorb the laser energy which is called inverse Bremsstrahlung. If the pulse duration time is shorter than both the electron cooling time and the time scale for energy transfer to the lattice which is at the order of 1 picosecond [7,8], molten phenomenon will not happen and plasma shapes as an intermediate state in the hole [9]. So femtosecond laser machining is a kind of cold processing.It has achieved more and more applications in the precision drilling field and some special circumstances in recent years, for example, drilling in the new generation of single crystal engine blades or engine blades with ceramic coating which are not conductive.

Although femtosecond laser machining has shown its advantages in precision drilling, it is very difficult to evaluate the quality of the hole in situ and real-time, which is of great usage to guide processing. This defection comes from the lack of diagnostics for hole parameters. The hole depth is one of the most important parameters during hole drilling. In some circumstances, blind holes with certain depths are required as is shown in figure 1(a) and the operators need to know the real-time hole depth to judge at what time to switch off the machining laser. In some other cases, although though holes are required the machining laser should also be switched off timely at the moment of penetration, avoiding unwanted hurt onto the components behind the though holes which is called 'backside damage' as is shown in figure 1(b). This measurement need is expressed fully in air-film hole drilling for engine blades which have double-layer structure and for engine combustion chambers which are annular. Optical cameras are used as the main tools in most of the laser drilling lathes to observe the hole depth. One of the two typical viewing angles is located to take photos of the hole as is illustrated by camera 1 or 2 in figure 1 (b). The shortcomings of this kind of direct observation are obvious. Firstly, during drilling the plasma in the hole will give out a harsh light which will obstruct the camera from real-time judging of the hole depth, so this method is not real-time. Secondly, this direct optical diagnostic is unable to measure quantitatively and it can only works as a reference. Thirdly, the metal to be processed can not give out light and the photos in the small holes are always too fuzzy to be distinguished. Optical coherence tomography (OCT) [10], which is widely used in the field of two-dimensional topography of human issues [11], is potentially applicable in hole drilling in metal materials. The inline coherent imaging (ICI) diagnostic has been proposed by P.J.L. Webster [12] which is capable of guiding blind hole percussion drilling basing a similar manner to spectral domain optical coherence tomography (SD-OCT) [13] and it has been proven to be applicable in bone [14] and ceramic [15] etching. In these experimental attempts, picosecond lasers or nanosecond lasers are used as the machining light sources which means the laser pulse is longer than the electron cooling time scale. So it is not a cold processing and melt pool exists in the hole. This dynamic is quite different to the physics of femtosecond processing in which plasma dominates the phase change during processing. The existence of plasma in the optical path of a coherent imaging diagnostic will obviously affect the path length and the phase of light, which are the right mechanisms of an interferometer [16] and a phase contrast imaging [17] diagnostic in plasmas, respectively.

In this article, a real-time depth diagnostic basing on in-process coherent imaging is presented, together with plasma density amendment. This diagnostic is now working as a tool to measure hole depth in air-film hole drilling in engine blades. The rest of this article is arranged as follows. Section 2 is a brief introduction of the physics of the coherent imaging diagnostic. The development of the coherent imaging diagnostic is given in section 3. The influence of plasma density to the optical path length is discussed in section 4. In section 5, there is a conclusion.

## 2. PRINCIPLE OF COHERENT IMAGING

The coherent imaging diagnostic presented in this article bases on the fundamental physics of light interference. It is in fact an upgraded spectral domain fiber Michelson interferometer. Broadband light is used as the diagnosing beam which is divided into two coherent parts. One has a fixed length of optical path which is call the reference arm, and the hole depth value is coupled into the other part which is called the sample arm and changes along with increasing of hole depth. The light beams reflected from these two arms will interfere in the output of the interferometer. Then the variation of depth can be easily extracted from the movement of the interference fringes.

In the femtosecond laser drilling, the machining laser is composed of a series of laser pulses $p$. These pulses interact with the material in the hole and shapes a set of hole depths $z_i|_{i=1 \to p}$ during drilling. These interacting material surfaces are considered to shape a set of $p$ interfaces in the sample arm each with a progressive optical path length difference of $2z_i|_{i=1 \to p}$ to the optical path length of reference arm. The intensity of the spectral interference fringe is then can be expressed as [13,18]

$$I(k, i) = A(k) \left[ \frac{I_{\text{Ref}}}{2p} + \eta \frac{I_i}{2} + \sqrt{\frac{\eta}{p} I_{\text{Ref}} I_i} \cos(2kz_i)|_{i=1 \sim p} \right], \tag{1}$$

where $k$ stands for the wavenumber, $A(k)$ stands for spectral envelope of the light, $I_{\text{Ref}}$ stands for the light intensity from the reference arm, $\eta$ stands for the reflectivity of the interacting material surface and $\eta I_i$ stands for the light intensity reflected by the $i^{\text{th}}$ interacting material surface. The first term in the right hand is a constant and it can be recognized by means of writing a tailored code. The second term is very small comparing with the first term because normally $\eta \ll 1$ and this term can be neglected. The third is the interference term which cosine varies and can be measured by a linear detector array. The depth information is hidden in this term and it can be extracted from the detected interference fringes via Fast Fourier Transform. The depth-reflectivity evolution during drilling can be obtained and the depth appears as a point spread function. The center of the function is the depth and the full width at half maximum (FWHM) is defined as the resolution $\Delta z_{\text{FWHM}}$ of depth. For a Guassian spectrum,

$$\Delta z_{\text{FWHM}} = \frac{2\ln 2}{\pi} \frac{\bar{\lambda}^2}{\Delta \lambda_{\text{FWHM}}}, \qquad (2)$$

where $\bar{\lambda}$ stands for the central wavelength for a Gaussian spectrum and $\Delta \lambda_{\text{FWHM}}$ stands for the FWHM of the broadband wavelength.

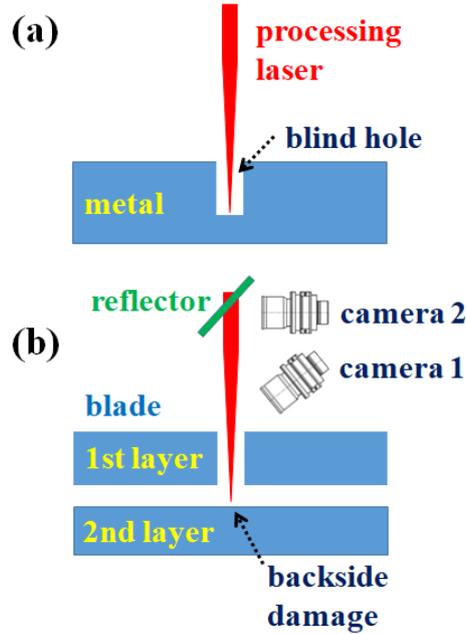

Figure 1. Two typical circumstances in which the hole depth diagnostic is potentially useful: (a) blind hole drilling; (b) hole drilling in a double-layer component. Camera 1 and 2 indicate two typical styles of direct hole viewing by using cameras.

## 3. DEVELOPMENT OF DIAGNOSTIC

The diagram of this real-time coherent imaging diagnostic is shown in figure 2, which borrows the idea of a spectral domain Michelson interferometer. It consists of five parts, i.e. the coherent light source, the reference arm, the sample arm, the detection system and data processing system. A super-luminescent diode with a central wavelength of $\bar{\lambda} = 830$ nm and a FWHM of $\Delta\lambda_{\text{FWHM}} = \pm\, 13$ nm is used as the coherent light source. According to equation (2), the depth resolution is $\Delta z_{\text{FWHM}} \approx 12\,\mu\text{m}$. The light beam is then split equally to two sub-beams by a four-channel splitter. One sub-beam is led into the reference arm and the other is led into the sample arm. The reference sub-beam is reflected by a gold-coated mirror with a fixed length of the optical path. The sample sub-beam exits from a collimating lens and then is reflected by a dichroic mirror where this diagnostic beam is combined with the machining laser. The combined beam is focused by a lens group. A 1-mm-thickness blade alloy sample to be machined is located with its surface perpendicular to the beam and the beam focus is set right on the sample surface. A part of the diagnosing sub-beam will be reflected back into the collimating lens with a reflectivity $\eta$. When the reflected sample sub-beam and reference sub-beam meet in the splitter, interference phenomenon happens. By using a volume grating, spectral domain interference fringes shape which can be captured and recorded by a line camera. Then these fringe data are sent to a computer and the third term in the right side of equation (1) can be identified by the code which has been developed by the authors. During drilling, as the hole depth varies continuously, the fringes in the line camera vary accordingly. By means of integrating a suitable algorithm in the code, the depth information can be extracted. It is to be noted here that, in the optical path of this coherent diagnostic, single-mode fibers are used as the main medium for coherent light transmission. These fibersare parent to the light which has a wavelength of around 830 nm to filter out all other background lines, such as the reflected machining beam and emission from plasma.

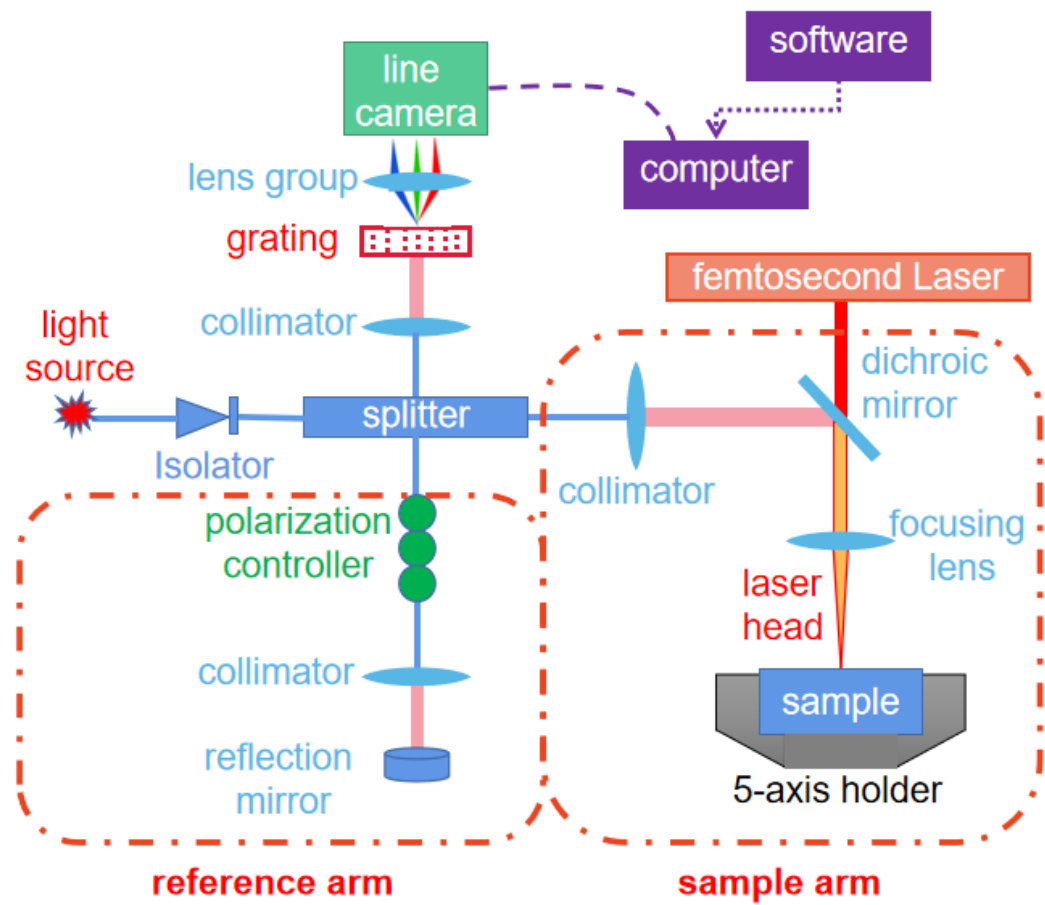

Figure 2. Diagram of the coherent imaging diagnostic.

**4. PLASMA AMENDMENT**

As has been mentioned above, the flamed metal material will generate plasma in the hole when a femtosecond laser pulse interacts with sample material. The existence of plasma indicates a desired cold processing which not only increases the machining efficiency but also improves the hole quality. However, the plasma also brings disadvantages. The most detrimental effect is that the plasma will deposit back onto the inner hole wall and shapes recast layer if the plasma can not be pulled out of the hole immediately. If the hole is deep, the recast layer material can absorb the energy from the next laser pulse and will re-generate plasma in the hole. This 'plasma- recast layer- plasma ' cycle repeats itself resulting in an extremely low efficiency during deep hole drilling. Exploring plasma property is of great importance for finding a way to solve this problem but this is out of the scope of this article and will be discussed elsewhere. For the diagnostic presented in this article, plasma in the hole will cause the variation of optical path length comparing with an optical path in the air. The plasma density is the key parameter determining the optical thickness of plasma.

For a coherent imaging diagnostic which is delicate, plasma influence to the optical path length is maybe non-ignorable and plasma amendment should be carefully investigated even if the volume of plasma is very small. Figure 3 shows the experimental set up of plasma density measurement during laser drilling. A femtosecond fiber laser which emits 400 fs pulses at a controllable frequency of 50-200 kHz with a maximal single-pulse energy of 50 μJ is used as the machining laser.

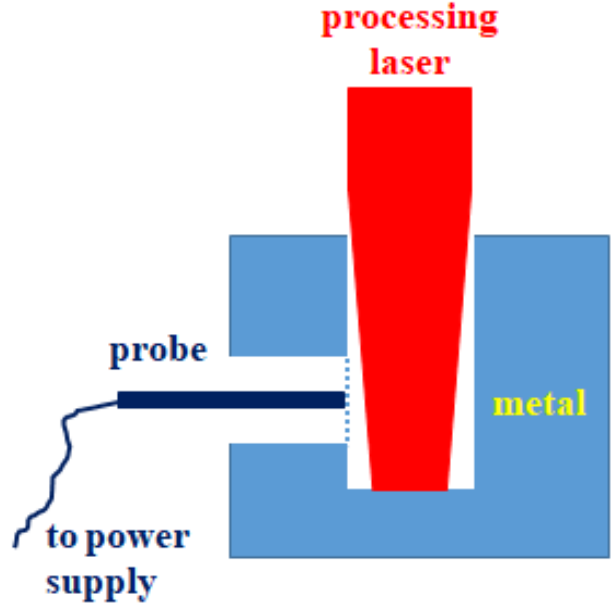

Figure 3. Experimental set up of plasma density measuring during laser drilling.

A sample which is made of the same alloy to the engine blade has been machined early as is shown in figure 3 before the experiment. A perpendicular cylinder hole with a cross diameter of about 1mm is prior prepared. Another hole with the same diameter is machined from the left side of the sample, which penetrates the sample and shapes a side opening in the wall of the perpendicular hole. A tungsten wire with a radii of $r = 0.1$ mm is inserted into the side hole, working as a Langmuir probe tip. It is to be noted here that the probe tip is located just right behind the opening, hidden in the shadow of the hole wall from touching the machining laser beam. This spatial limitation is marked by the blue dot line in figure 3. The probe is linked to a scanning power supply. During drilling, this probe can provide the plasma density information $n_e$ [19]. Traditional Langmuir probe is to measure plasma parameters in a vacuum vessel, which is of great difference to what we are talking about in the air. Particle collision effect should be considered here. We use following model to modify the measured plasma density $n_{e,\text{mod}} = \frac{1}{K} n_e$, where $K \equiv 1 - \frac{1}{2}\left[1 - \frac{1+\alpha}{(1+2\alpha)^{1/2}} + \right.$

$\frac{\alpha^2}{(1+2\alpha)^{1/2}(1+\alpha)}]$ stands for the coefficient of correction and $\alpha \equiv r/\zeta$ stands for the ratio of probe radii $r$ to the mean free path $\zeta$ which is determined by the air pressure.

The dispersion relation of electromagnetic wave in plasma is $\omega^2 = c^2 k^2 + \omega_{pe}^2$, where $\omega_{pe} = \left(\frac{4\pi n_{e,mod} e^2}{m_e}\right)^{1/2}$ stands for electron angular frequency, c stands for light speed and $m_e$ stands for the electron mass. The phase velocity of the wave is $v_p = \frac{\omega}{k} = \sqrt{c^2 + \frac{\omega_{pe}^2}{k^2}}$. Then the refractive index of plasma can be calculated by $= \frac{c}{v_p} = \sqrt{1 - \frac{\omega_{pe}^2}{\omega^2}}$.

The measured plasma density in the hole is a local value of about $6.9 \times 10^{21}/m^3$. It is very difficult to get the vertical profile of plasma density in the hole. Given the pressure balance in the hole, a simple approximation is regarding the plasma in the hole as uniformity. Substituting all the constants and parameters into above equations, the optical path length amendment value $\Delta k$ because of the existence of plasma in the hole is shown in figure 4, which is a linear relation and can be expressed as $\Delta k = 0.349d$, where $d$ stands for the instantaneous value of hole depth in millimeter and $\Delta k$ is in wavenumber. This result indicates that when the depth value reaches $\frac{1}{0.349} \approx 2.9$ mm, the optical path length error is up to one wavelength, i.e. 830 nm. Typically, the depth of vertical air-film holes in engine blades is at the order of 1~2 mm, the maximal measurement error caused by plasma is less than 1 μm. This error is ignorable comparing to the resolution $\Delta z_{FWHM}$ of depth. While in some other circumstance, such as tilt hole drilling in the engine blades or combustions which may have depth of over 10 mm, this results in a measurement error of about 3μm or even larger. This error is no longer ignorable and amendment should be done.

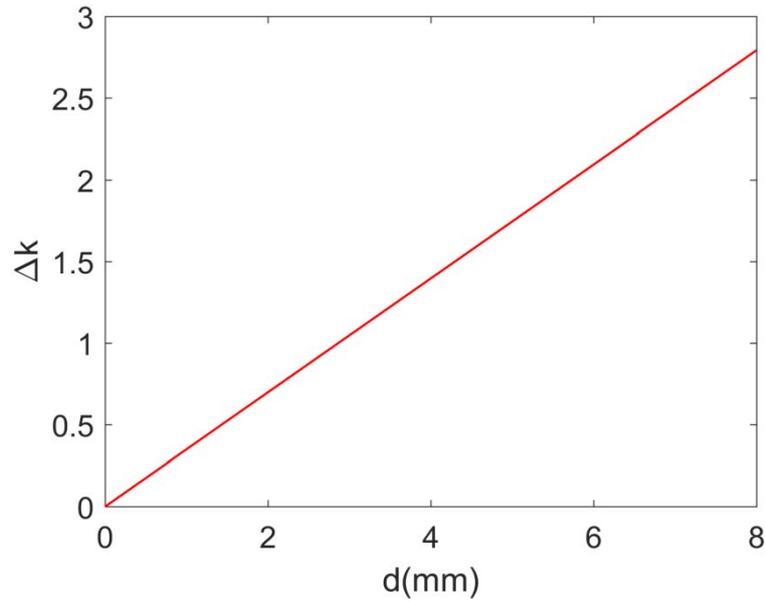

Figure 4. The optical path length amendment value (unit : number of wavelength) to the hole depth (unit : mm).

## 5. EXPERIMENTAL RESULTS

This coherent imaging system has been installed as the real-time inline depth diagnostic in real femtosecond laser machining lathes for engine blade hole drilling. One typical machining shot is discussed here to show the performance of this coherent imaging diagnostic. In this shot, the thickness of the blade material sample is about 1 mm. The interference fringes before drilling recorded by the line camera and the coherent imaging code are shown in figure 5(a).The x-coordinate is the pixel of the line camera which has 2048 pixels. At a certain time point, the coherent fringes are pictured and recorded as such an array which has 2048 values. The code can show last 20 arrays arranged along y-coordinate. In this experiment, the sampling rate of the camera is set to be 20 kHz which means the time difference between nearby fringe arrays is 50 μs. These fringe arrays are partially zoomed in as is shown in figure 5(b). By extracting the depth value by referring to equation (1) and using Inverse Fourier Transform, the real-time depth diagnosing is realized. Figure 6 shows the evolution of the hole depth during drilling with plasma amendment which confirms the realization of the coherent imaging diagnostic. Except for real-time depth value, lots of other useful information can be read out from this figure. For example, the slope of the curve indicates the drilling speed and the characteristic of locally non-monotonic decreasing implies a rugged surface in the hole.

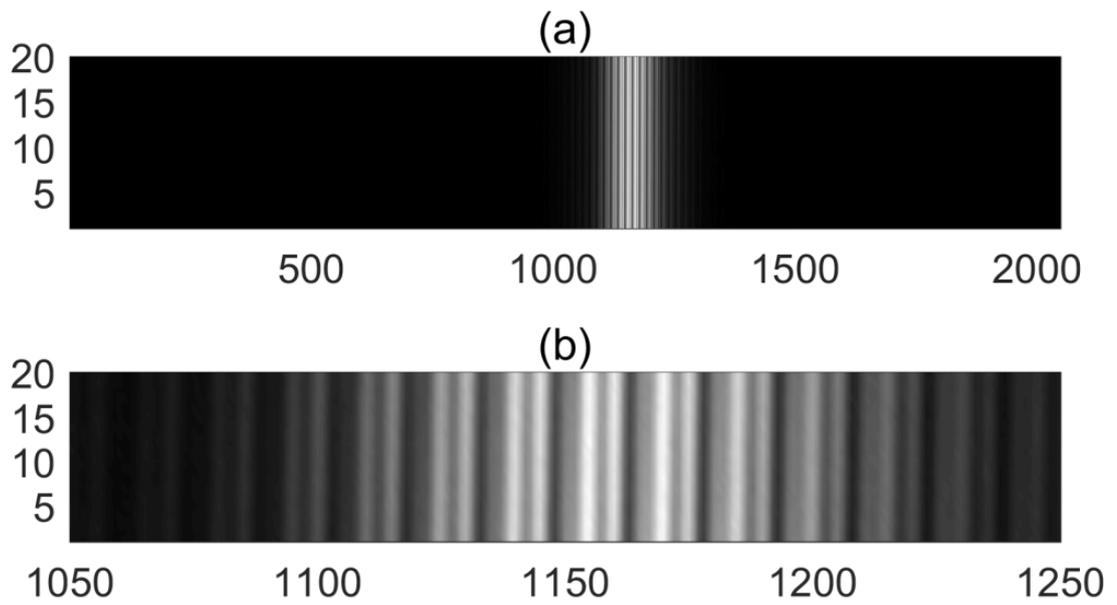

Figure 5. (a) Interference fringes recorded by the 2048-pixel line camera and the coherent imaging code. (b) Partial enlarged interference fringes.

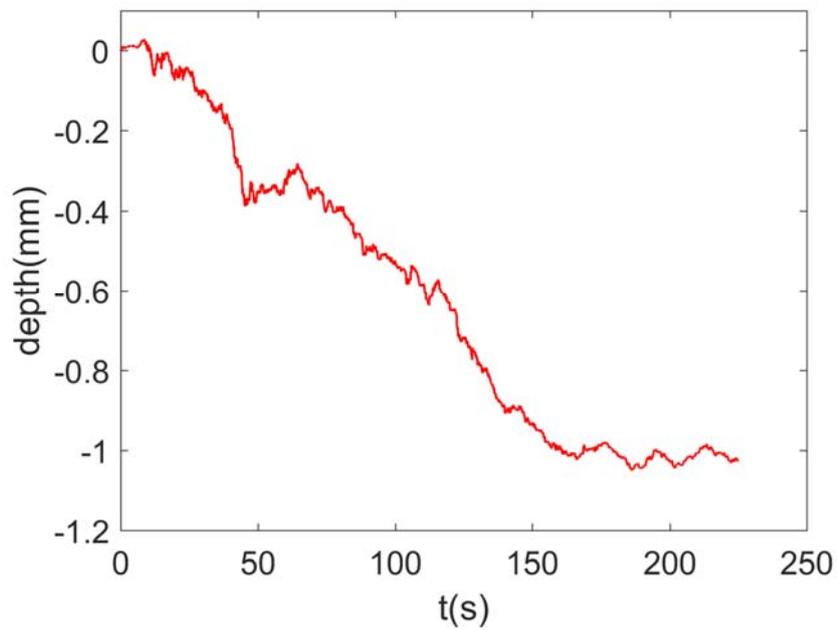

Figure 6. The real-time measured hole depth during drilling in an engine blade.

## 6. CONCLUSION AND DISCUSSION

A real-time depth diagnostic based on in situ coherent imaging has been developed for femtosecond laser drilling. A code is also written to extract the evolution of hole depth. For current state, the real-time depth is calculated and displayed as a reference to guide the researchers. The fast background data processing to produce a reliable trigger signal to switch on/off the machining laser timely is much more important than a depth display. This is out of the scope of this article and will be a good future work.

Plasma effect is considered carefully to amend the real-time depth value based on experimentally measured plasma density. But, remarkably, this measurement is under a special drilling condition, for example, a certain machining laser and sample material to be drilled. Although this particularity should be considered to evaluate the merit and application range of this diagnostic, it is not important in real machining industry based on two points. Firstly, the work presented in this article confirms the physical and technical feasibilities of this coherent imaging method in hole drilling in engine blades. Secondly, common practice is to prepare tens of or more lathes to machine one kind of blade. The lathe should be designed based on above investigation to meet this particularity and the application range is meaningless here. If one want to drill holes with some other requirements, another tailored coherent imaging diagnostic with its code should be developed to match the new lathe model.

## ACKNOWLEDGEMENTS

The authors thank Prof. H.J. Ren from University of Science and Technology of China for the great help of experiments and code writing. This work was partly supported by National Key R&D Program of China under Grant No. 2022YFE03100002.